# Embedding information in physically generated random bit sequences while maintaining certified randomness

Shira Sardi[1,(a)], Herut Uzan[1,(a)], Shiri Otmazgin[1], Yaara Aviad[3,4], Michael Rosenbluh[1,3] and Ido Kanter[1,2,(b)]

[1]Department of Physics, Bar-Ilan University, Ramat-Gan 52900, Israel.
[2]Gonda Interdisciplinary Brain Research Center, and the Goodman Faculty of Life Sciences, Bar-Ilan University, Ramat-Gan 52900, Israel.
[3]Department of Physics, The Jack and Pearl Resnick Institute for Advanced Technology, Bar-Ilan University, Ramat-Gan, 52900 Israel.
[4]Department of Applied Mathematics, Jerusalem College of Technology (JCT), Jerusalem, Israel.
[(a)]These authors contributed equally to this work.
[(b)]E-mail: ido.kanter@biu.ac.il

Ultrafast physical random bit generation at hundreds of Gb/s rates, with verified randomness, is a crucial ingredient in secure communication and have recently emerged using optics based physical systems. Here we examine the inverse problem and measure the ratio of information bits that can be systematically embedded in a random bit sequence without degrading its certified randomness. These ratios exceed 0.01 in experimentally obtained long random bit sequences. Based on these findings we propose a high-capacity private-key cryptosystem with a finite key length, where the existence as well as the content of the communication is concealed in the random sequence. Our results call for a rethinking of the current quantitative definition of practical classical randomness as well as the measure of randomness generated by quantum methods, which have to include bounds using the proposed inverse information embedding method.

**Introduction.** – Emerging classical and quantum secure communication methods rely on long, truly random, bit strings, which are used to scramble the information by applying mathematical functions. The mathematical definition of an ideal process for generating random bit sequences is straightforward - the next bit in the sequence is zero or one with equal probability independent of the previous bits. However, from a practical viewpoint, given a seemingly random sequence, how can one be sure that it is indeed random? It is possible that such a question is impossible to answer. Under such circumstances what does it imply about secure communications?

Two decades ago the only available fast random bit generators (RBGs) were based on pseudo-random generators[1, 2]. The security using such random bit sequences is, however, undermined as a result of the deterministic nature of their generation. Concern about security has become even greater with the emergence of more sophisticated attacks on the growing communication networks with ever-increasing numbers of end users. The emergence of non-deterministic ultrafast physical RBGs began in 2008, based on the combined threshold and XOR gated bit outputs of two chaotic diode lasers, producing 1.7 Gb/s random bits with certified randomness[3]. The physical implementation required some constraints such as an incommensurate ratio between the lengths of the external laser cavities of the two similar lasers. A method based on a single chaotic laser whose chaotic intensity fluctuations are digitized and multi-bit extracted (fig. 1) was introduced in 2009 and produced 12.5 Gb/s[4]. Subsequently, the single laser method was generalized and achieved generation rates of 300 Gb/s with certified randomness using higher order derivatives of the digitized signal[5]. This method has been further adopted and generalized[6, 7] with rates moving toward Tb/s[8, 9] using many variants including different types of lasers[10-

16], an implementation on integrated circuits[7, 17] and electronic chaotic elements[18].

The verification of the randomness produced by any physical RBGs[19] or pseudorandom bit generators is binary quantified by several statistical test suites[20-22]. The most popular one is the NIST test suite[22] consisting of many local and global tests and typically requires exhaustive examination of a very long random bit sequence of at least 1 Gb. Successfully passing one or several of these statistical test suites has become the standard certification of acceptable randomness of the sequence being tested and of the method used to generate the sequence. These statistical test suites, however, only provide a fail or pass criterion for the random bit sequence and do not give a quantitative score for the strength of the randomness in the sequence.

**Results.** – We propose a reverse strategy which aims to quantify the maximal amount of information that can be systematically embedded in a certified random bit sequence using a given test suite, without harming its certification. Using such an inverse strategy, we can quantify the level of randomness and go beyond a binary certification. In addition, since the information is systematically embedded in the bit sequence, this method offers a new type of classified cryptosystem, which is similar to sub-thermal communication. For simplicity, the certification of the randomness of a bit sequence used in this paper is based on the popular NIST test suite only.

The demonstration of the proposed method begins with the generation of a 1 Gb long random bit sequence which successfully passes the NIST test. The data is generated from the digitized (8 bit analogue to digital converter) output intensity fluctuations of a chaotic semiconductor laser using the 3rd derivative of the obtained data and retaining the 5 least significant bits of the derivative value5 (fig. 1). The generated sequence is tested by the NIST test suite and thus certified to be random. The 1 Gb long sequence is then partitioned into 1000 blocks of 1 Mb each. In the second step, each block of 1 Mb is manipulated by dividing it into successive multiple segments of K+G bits (fig. 2(a)). The information is embedded in the last bit (colored red in fig. 2(b)-(c)) of each segment K (orange), where the region G (blue) indicates the gap between successive segments of length K. In the first embedding scheme we examined, the last bit of each segment K is set to the parity of its previous K-1 bits (fig. 2(b))

$$X^S_{m+K-1} = mod_2\left(\sum_{p=m}^{m+K-2} X_p\right) \quad (1)$$

where $X^S_{m+K-1}$ represents the last bit of each segment of length K (fig. 1(a)), after the modification of Eq. (1). In a second embedding scheme we examined, $X^S_{m+K-1}$ is set to the parity of a subset of its preceding K-1 bits. The limiting case consisting of a subset of only two bits $X_m$ and $X_{m+q}$ resulting in

$$X^S_{m+K-1} = mod_2(X_m + X_{m+q}) \quad (2)$$

where m and q are arbitrary integers with a value between 1 and K-1 (fig. 2(c)). The implementation of either of these schemes indicates that the last bit of each orange-segment of K-bits (fig. 2) is not independent and can be uniquely deduced from the preceding K-1 bits.

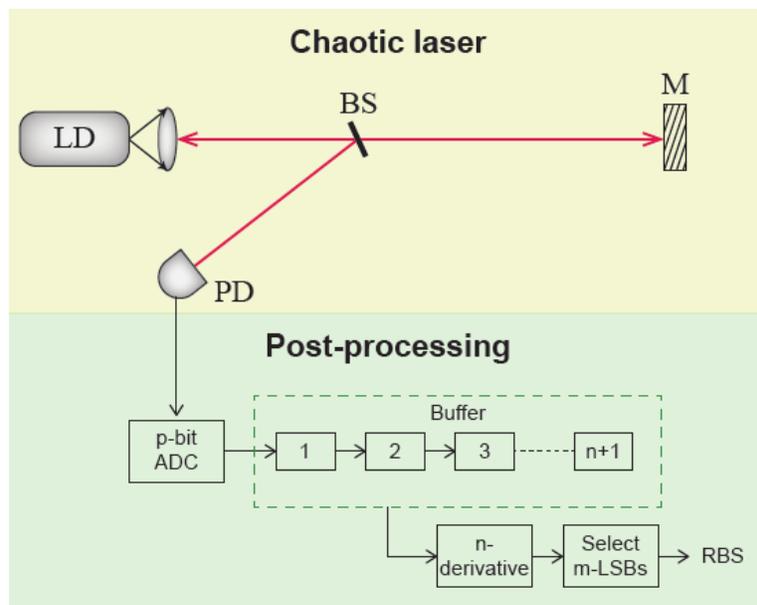

Fig. 1: (color online). Schematic of a physical RBG. The optical setup (yellow) consists of a LD, laser diode; M, mirror; BS, beam splitter and PD, high-speed photodetector. The post-processing (green) consists of p-bit ADC followed by the nth derivative of the digitized signal and finally the appending of the m-least-significant-bits (LSBs) to the random bit sequence.

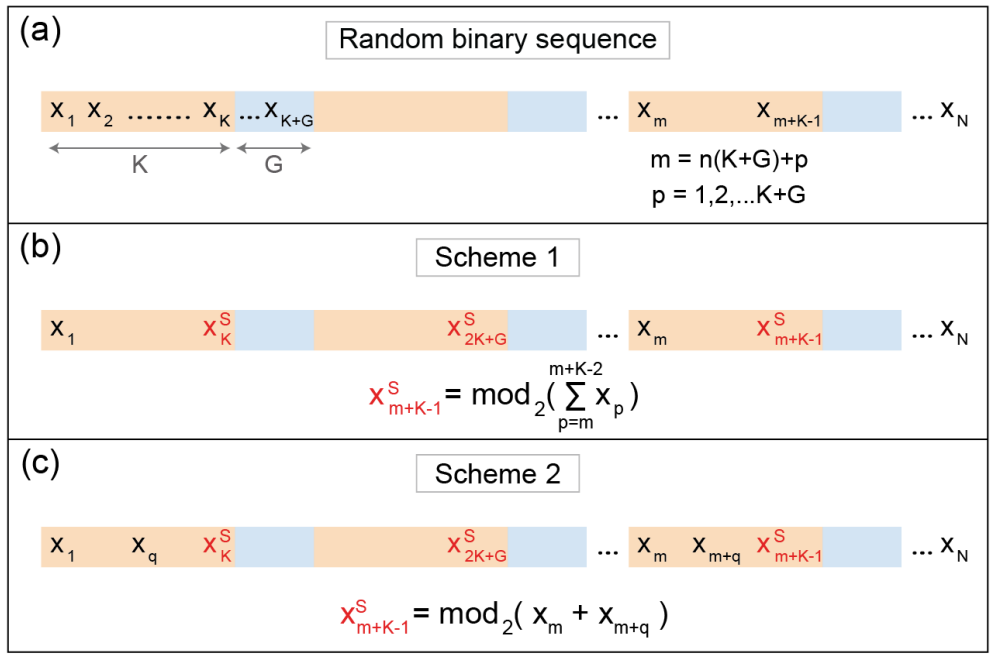

Fig. 2: (color online). The inverse method whereby information is embedded in a random bit sequence. (a) Each random binary block of length N (1 Mb in our data), is divided into segments of length K+G, a segment of length K (orange) followed by a gap segment of length G (blue) and n denotes the segment number. (b) The first scheme for embedding information, where the last bit of each K-segment equals the parity of its first K-1 bits. (c) The second embedding scheme where the last bit of each K-segment equals the parity of two given previous bits belonging to the same K-segment.

|  | Original sequence | | Scheme 1<br>K = 55, G = 5 | | Scheme 2<br>K = 59, G = 5<br>m = 1, q = 23 | | Scheme 1<br>K = 20, G = 5 | | Scheme 2<br>K = 19, G = 5<br>m = 1, q = 7 | |
|---|---|---|---|---|---|---|---|---|---|---|
| Statistical test | P value | Proportion | P value | Proportion | P value | Proportion | P value | Proportion | P value | Proportion |
| Frequency | 0.6642 | 0.99 | 0.5852 | 0.99 | 0.7459 | 0.989 | 0.1069 | 0.988 | 0.5626 | 0.988 |
| Block Frequency | 0.8092 | 0.99 | 0.4983 | 0.994 | 0.9453 | 0.991 | 0.3221 | 0.991 | 0.7887 | 0.99 |
| Cumulative Sums | 0.377 | 0.99 | 0.6973 | 0.988 | 0.7656 | 0.988 | 0.8862 | 0.985 | 0.1719 | 0.987 |
| Runs | 0.2826 | 0.99 | 0.5503 | 0.992 | 0.8862 | 0.989 | 0.9602 | 0.991 | 0.7656 | 0.992 |
| Longest Run | 0.3925 | 0.993 | 0.8564 | 0.987 | 0.2609 | 0.986 | 0.5811 | 0.988 | 0.2757 | 0.993 |
| Rank | 0.8817 | 0.985 | 0.0757 | 0.989 | 0.5503 | 0.989 | 0.3977 | 0.984 | 0.2144 | 0.99 |
| FFT | 0.0448 | 0.989 | 0.3736 | 0.992 | 0.5362 | 0.989 | 0.1259 | 0.989 | 0.2296 | 0.985 |
| Non-overlapping Template | 0.475 | 0.983 | 0.4391 | 0.983 | 0.7177 | 0.982 | 0.7238 | 0.982 | 0.6849 | 0.983 |
| Overlapping Template | 0.6993 | 0.995 | 0.9987 | 0.983 | 0.7116 | 0.989 | 0.5003 | 0.989 | 0.5955 | 0.991 |
| Universal | 0.4616 | 0.99 | 0.8201 | 0.989 | 0.5728 | 0.991 | 0.7734 | 0.989 | 0.0529 | 0.99 |
| Approximate Entropy | 0.5042 | 0.99 | 0.2405 | 0.99 | 0.6725 | 0.991 | 0.2044 | 0.989 | 0.0767 | 0.991 |
| Random Excursions | 0.4373 | 0.9826 | 0.767 | 0.9792 | 0.1027 | 0.9844 | 0.2472 | 0.9868 | 0.6355 | 0.9797 |
| Random Excursions Variant | 0.458 | 0.9826 | 0.4691 | 0.9824 | 0.6793 | 0.9859 | 0.3319 | 0.9801 | 0.9104 | 0.9875 |
| Serial | 0.939 | 0.987 | 0.5565 | 0.984 | 0.1512 | 0.994 | 0.2442 | 0.988 | 0.5483 | 0.985 |
| Linear Complexity | 0.6496 | 0.989 | 0.7157 | 0.994 | 0.892 | 0.989 | 0.3753 | 0.988 | 0.9432 | 0.987 |
| Result | Success | | Success | | Success | | Success | | Success | |

Fig. 3: (color online). Results of the NIST test. Results of NIST Special Publication 800-22 statistical tests, using 1000 samples of 1 Mb data and significance level $\alpha$ = 0.01. A "success" result requires that the P value (uniformity of P values) should be larger than 0.0001 and the proportion should be greater than 0.980561. Successful test results are shown (left to right): the original sequence; Embedding scheme 1 with K=55 and G=5 (fig. 2(b)); Embedding scheme 2 with K=59, G=5, m=1 and q=23 (fig. 2(c)); Embedding scheme 1 with K=20 and G=5 (and 42 excluded blocks); Embedding scheme 2 with K=19, G=5, m=1 and q=7 (and 20 excluded blocks). All embedding schemes where implemented on the same original sequence.

The implementation of embedding scheme 1 (fig. 2(b)) on the original sequence (which passed the NIST test as shown in the first two columns of fig. 3) with K=55 and G=5, for instance, results in a new random bit sequence, which also successfully passes the NIST test. This is also the case for embedding scheme 2 (fig. 2(c)) with K=59, m=1, q=23 and G=5. Successfully passing the NIST test is robust to the value of K, typically > 15 (see appendix fig. A1) and for positive and negative G values, with the negative values indicating partially overlapping K-segments (see appendix fig. A2). Note that the transition between failure/success at small/large K is not sharp (fig. A1) as a result of local fluctuations along the random sequence. For some of the examined K values a failure of the NIST test was obtained where typically one of the battery of NIST tests was slightly below the expected threshold (proportion = 0.980561). In such events, we first identified the 1 Mb blocks responsible for the failure of the specific test for which the threshold was not reached, using the detailed NIST report. The number of such blocks is typically in the range of 20 to 40, which is small compared to the total of 1000 blocks (fig. 3). Next, we implemented the embedding scheme on the original sequence without modifying the few blocks that failed, and this results in a random bit sequence which successfully passes the NIST test (fig. 3 and see also appendix fig. A1).

The fraction of potentially modified bits as a consequence of the realization of the embedding schemes exceeds 0.01 (K ~ 100) and can even come close to 0.1 (K → 10) (fig. 3 and see also appendix figs. A1 and A3). Note that the proportion of the NIST test is not monotonically increasing with K and can even move away from the threshold by the implementation of the scheme. Though these bits can be uniquely extracted from the rest of the unmodified bits, the randomness of the entire bit sequence is still verified by the NIST test. Thus, to an outside observer the embedded bit sequence looks statistically identical to the original random bit sequence in spite of the fact that it contains a significant amount of information. The robustness of successfully passing the NIST test was verified for at least eight other random bit sequences of 1 Gb each, as well as for varying G values. In addition, results were found to be robust to the generalization of Eq. (1) to a random embedding assignment

$$X^S_{m+K-1} = mod_2\left(\sum_{p=m}^{m+K-2} X_p + rand\_bit\right) \quad (3)$$

where $rand\_bit$ = 0/1 with equal probability, e.g. compressed data, and similarly for the second embedding scheme, Eq. (2) (see appendix fig. A4). Similar results were also obtained for biased statistical prescriptions for $rand\_bit$, e.g. uncompressed data, as well as original random bit sequences generated by a pseudorandom number generator (not shown).

The embedding scheme which determines the last bit in each orange-segment of K-bits can be generalized for any subset of the K-1 bits (see for instance appendix fig. A5), hence the number of possible schemes grows exponentially with K

$$No. of\ embedding\ schemes = 2^{K-1} \quad (4)$$

In principle this number can be much enlarged using predefined varying K values and schemes along the sequence. Alternatively, one can use a fixed K value and predefined varying q values, e.g. eq. (2), along the sequence, which found in some cases to pass the NIST test (not shown). This space of possible schemes serves as a huge key-space for the proposed private-key cryptosystem (fig. 4), in which a predefined key of length O(1), independent of the size of the message, is used at the encoder to cipher a binary message (fig. 4(a)). The decoder retrieves the message by reversing the embedding scheme on the received message (fig. 4(b)). An eavesdropper, who has full access to the transmitted signal, cannot distinguish between a random transmitted binary signal, containing only random noise, and the ciphered one. Both transmitted signals are interpreted as random sequences which successfully pass the NIST test. An exhaustive search for the concealed message faces the difficulty of guessing the embedding key from the huge key-space, Eq. (4).

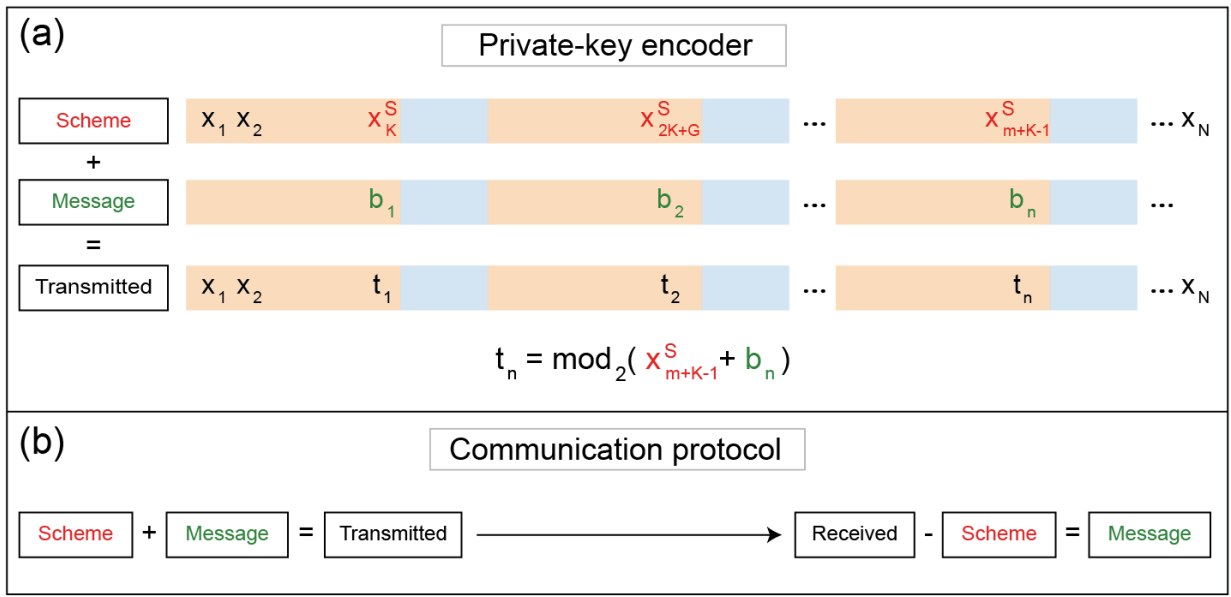

Fig. 4: (color online). Schematic of a type of sub-thermal classified private-key cryptosystem, where the existence of the communication itself is concealed. (a) A schematic private-key encoder. The message (green) is added to the embedding scheme (red), as exemplified in fig. 2(b)-(c), in order to create the transmitted sequence. (b) A schematic communication protocol using a private-key, selected from the huge space of schemes.

**Conclusion.** – The proposed private-key method has the unique feature similar to the class of sub-thermal communication or steganography[23], where the existence of a meaningful communication is concealed[24]. This goal is achieved using a communication channel[25] that is continuously occupied by the transmission of random bits at arbitrary transmission rates, which are embedded from time to time by a message using an agreed upon embedding scheme, such as Eqs. (1)-(2). An observer cannot distinguish the information containing sequence and the random transmission, since both successfully pass the NIST test suite. The capacity of such classified communication is large, since the ratio can exceed 0.01 as $K \lesssim 100$ (fig. 3). In the limit where the rate approaches $O(0.1)$, $K$ below fifty (fig. 3), the embedding scheme has to incorporate variation along the sequence in order to achieve a large enough key-space. This allows secure communication to be maintained even against an exhaustive search attack for the embedding scheme used. In cases where the information embedded sequence fails the statistical test as a result of the failure of several blocks, one possible practical solution is to delete the failing blocks and append an equal number of random blocks to the end of the original random bit stream.

The suggested inverse method and its utilization in secure communication is also applicable to test the stability of quantum random bit generators (QRBGs) against injection of information. The generation rates of QRBGs are typically several order of magnitudes below Gb/s[26, 27], unless a classical post-processing or additional classical sub-processes are concatenated to the original pure quantum fundamental source[28, 29]. In QRBGs the quality of the randomness is assumed to be perfect, as a consequence of the fundamental principles of quantum physics. In practice, this perfect randomness is possibly diminished by many experimental imperfections, such as detector quantum efficiency, for example. Hence, a sequence generated by a QRBG also has to be certified by statistical test suites[26-29]. Moreover, even assuming that there are no experimental imperfections and that a QRBG is indeed producing ideal random bit sequences, the utilization of these sequences for practical use (e.g. fig. 4) requires a certification of their randomness. Therefore, an effective statistical test, which can distinguish between an original quantum guaranteed sequence and spurious sequences, is necessary.

Our method suggests that one can test the strength of the randomness in a string by systematically and gradually adding information to a random sequence generated either by a physical random source or by a QRBG, until the information containing sequence no longer passes the statistical tests, such as NIST. The minimal $K$ value at which this failure sets in (with a given probability, since the transition is not sharp) is suggested as a good indicator of the strength of the randomness. It is intriguing to consider whether there are other statistical tests that could yield a quantitative measure of the amount of information that can be embedded in certified random bit sequences and whether this quantity differentiates quantum from classical generated sequences.

**Appendix: additional data.** –

(a)

| K value | 9 | 10 | 11 | 13 | 20 | 23 | 27 | 36 | 45 | 54 | 63 | 72 | 81 | 90 | 99 | 108 |
|---|---|---|---|---|---|---|---|---|---|---|---|---|---|---|---|---|
| Test Result | F | F | F | F | S | S | F | S | S | S | S | S | S | S | S | S |
| Number of removed blocks | 0 | 0 | 0 | 0 | 42 | 0 | 0 | 0 | 22 | 0 | 0 | 0 | 0 | 0 | 20 | 0 |

| K value | 116 | 125 | 134 | 143 | 152 | 161 | 170 | 179 | 188 | 197 | 206 | 215 | 224 | 233 | 242 | 256 |
|---|---|---|---|---|---|---|---|---|---|---|---|---|---|---|---|---|
| Test Result | S | S | S | S | S | S | S | S | S | S | S | S | S | S | S | S |
| Number of removed blocks | 0 | 34 | 0 | 39 | 0 | 15 | 20 | 0 | 20 | 15 | 0 | 0 | 14 | 0 | 21 | 0 |

(b)

| | Scheme 1 (sequence 2) $K=23$, $G=5$ | | Scheme 1 (sequence 2) $K=63$, $G=5$ | | Scheme 1 (sequence 2) $K=81$, $G=5$ | | Scheme 1 (sequence 2) $K=108$, $G=5$ | | Scheme 1 (sequence 2) $K=256$, $G=5$ | |
|---|---|---|---|---|---|---|---|---|---|---|
| Statistical test | P value | Proportion | P value | Proportion | P value | Proportion | P value | Proportion | P value | Proportion |
| Frequency | 0.9512 | 0.987 | 0.17 | 0.992 | 0.5422 | 0.992 | 0.3787 | 0.992 | 0.3994 | 0.99 |
| Block Frequency | 0.2983 | 0.991 | 0.2506 | 0.991 | 0.4983 | 0.993 | 0.7339 | 0.991 | 0.5708 | 0.991 |
| Cumulative Sums | 0.2225 | 0.987 | 0.4373 | 0.986 | 0.2676 | 0.99 | 0.7925 | 0.991 | 0.8463 | 0.99 |
| Runs | 0.6725 | 0.989 | 0.4136 | 0.988 | 0.0564 | 0.988 | 0.9493 | 0.992 | 0.2897 | 0.992 |
| Longest Run | 0.4136 | 0.988 | 0.3206 | 0.982 | 0.7963 | 0.991 | 0.3086 | 0.99 | 0.7379 | 0.991 |
| Rank | 0.0345 | 0.992 | 0.1776 | 0.991 | 0.9179 | 0.989 | 0.1107 | 0.992 | 0.8547 | 0.992 |
| FFT | 0.0624 | 0.984 | 0.0086 | 0.99 | 0.1495 | 0.993 | 0.5382 | 0.99 | 0.2743 | 0.988 |
| Non-overlapping Template | 0.1101 | 0.983 | 0.9153 | 0.983 | 0.2034 | 0.983 | 0.7715 | 0.982 | 0.389 | 0.981 |
| Overlapping Template | 0.8564 | 0.986 | 0.1209 | 0.989 | 0.0851 | 0.992 | 0.902 | 0.994 | 0.1767 | 0.99 |
| Universal | 0.8832 | 0.99 | 0.2248 | 0.988 | 0.9696 | 0.988 | 0.9357 | 0.993 | 0.5341 | 0.99 |
| Approximate Entropy | 0.9787 | 0.994 | 0.4012 | 0.99 | 0.7359 | 0.987 | 0.9216 | 0.994 | 0.6517 | 0.989 |
| Random Excursions | 0.7299 | 0.985 | 0.7826 | 0.9823 | 0.8257 | 0.9836 | 0.6187 | 0.9839 | 0.699 | 0.9856 |
| Random Excursions Variant | 0.2757 | 0.98 | 0.3925 | 0.9823 | 0.178 | 0.9869 | 0.3091 | 0.9855 | 0.095 | 0.984 |
| Serial | 0.333 | 0.99 | 0.0329 | 0.985 | 0.6683 | 0.99 | 0.0063 | 0.987 | 0.2609 | 0.988 |
| Linear Complexity | 0.0932 | 0.99 | 0.5852 | 0.99 | 0.4208 | 0.986 | 0.6163 | 0.985 | 0.4318 | 0.984 |
| Number of removed blocks | 0 | | 0 | | 0 | | 0 | | 0 | |
| Result | Success | | Success | | Success | | Success | | Success | |

Fig. A1: (color online). Results of the NIST test for different K values. Results of NIST Special Publication 800-22 statistical tests, using 1000 samples of 1 Mb data and significance level $\alpha = 0.01$. A "success" result requires that the P value (uniformity of P values) should be larger than 0.0001 and the proportion should be greater than 0.980561 (for Random Excursions and Random Excursions Variant tests the proportion value might be even lower). (a) Different K values for embedding scheme 1 with G = 5 (fig. 2(b)), implemented on the sequence from the manuscript (fig. 3), where F and S denote failure and success, respectively. (b) Successful test results for a different sequence than the one presented in the manuscript (fig. 3) are shown for embedding scheme 1 (fig. 2(b)) with G = 5 and (left to right): K = 23, 63, 81, 108, 256.

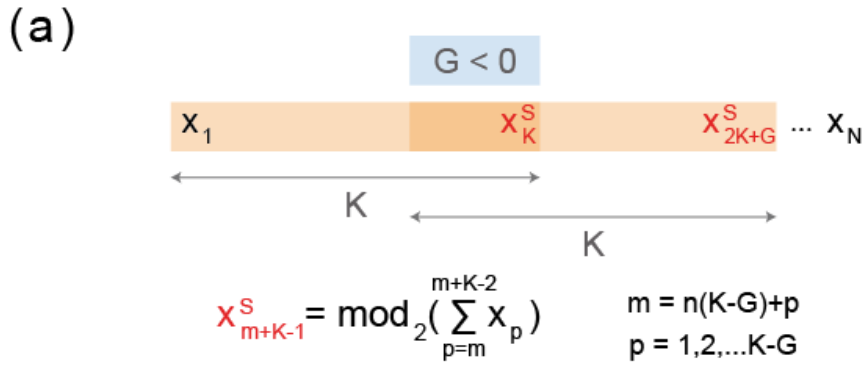

(a)

$$x^S_{m+K-1} = \mathrm{mod}_2\left(\sum_{p=m}^{m+K-2} x_p\right) \quad m = n(K-G)+p \quad p = 1,2,\ldots K-G$$

(b)

| Statistical test | Scheme 1 (sequence 1) K = 81, G = -7 | | Scheme 2 (sequence 3) K = 71, q = 29, G = -9 | |
|---|---|---|---|---|
| | P value | Proportion | P value | Proportion |
| Frequency | 0.092 | 0.991 | 0.4245 | 0.993 |
| Block Frequency | 0.1088 | 0.993 | 0.8891 | 0.991 |
| Cumulative Sums | 0.6475 | 0.989 | 0.4788 | 0.99 |
| Runs | 0.9549 | 0.991 | 0.2284 | 0.987 |
| Longest Run | 0.8645 | 0.99 | 0.4083 | 0.983 |
| Rank | 0.3703 | 0.989 | 0.4299 | 0.996 |
| FFT | 0.4559 | 0.989 | 0.4983 | 0.989 |
| Non-overlapping Template | 0.0712 | 0.981 | 0.8709 | 0.982 |
| Overlapping Template | 0.8786 | 0.994 | 0.1161 | 0.984 |
| Universal | 0.8935 | 0.999 | 0.7399 | 0.989 |
| Approximate Entropy | 0.2066 | 0.988 | 0.8429 | 0.992 |
| Random Excursions | 0.6059 | 0.9852 | 0.9476 | 0.9875 |
| Random Excursions Variant | 0.5681 | 0.9819 | 0.1257 | 0.9812 |
| Serial | 0.975 | 0.989 | 0.1767 | 0.993 |
| Linear Complexity | 0.8237 | 0.992 | 0.4208 | 0.988 |
| Number of removed blocks | 0 | | 0 | |
| Result | Success | | Success | |

Fig. A2: (color online). Results of the NIST test with overlapping segments, G < 0. (a) A random binary sequence divided into segments as in fig. 2(a). G < 0 resulting in an overlap between consecutive K segments. (b) Results of the NIST Special Publication 800-22 statistical tests, using 1000 samples of 1 Mb data experimentally obtained from the method presented in fig. 1. A "success" result requires that the P value (uniformity of P values) should be larger than 0.0001, a significance level $\alpha = 0.01$ and the proportion should be greater than 0.980561 (for Random Excursions and Random Excursions Variant tests the proportion value might be even lower). Successful test results are shown (left to right): Embedding scheme 1 with K = 81 and G = -7 (fig. 2(b)); Embedding scheme 2 for a different sequence than the one presented in the manuscript with K = 71, m = 1, q = 29 and G = - 9 (fig. 2(c)).

| Statistical test | Scheme 1 (sequence 1) K = 13, G = 5 | | Scheme 1 (sequence 1) K = 10, G = 5 | | Scheme 2 (sequence 1) K = 11, G = 5 m = 1, q = 6 | | Scheme 2 (sequence 1) K = 15, G = 5 m = 1, q = 7 | | Scheme 2 (sequence 2) K = 13, G = 5 m = 1, q = [2 6 9 12] | |
|---|---|---|---|---|---|---|---|---|---|---|
| | P value | Proportion | P value | Proportion | P value | Proportion | P value | Proportion | P value | Proportion |
| Frequency | 0.1564 | 0.993 | 0.1503 | 0.99 | 0.0686 | 0.987 | 0.0472 | 0.986 | 0.3489 | 0.992 |
| Block Frequency | 0.3457 | 0.993 | 0.829 | 0.991 | 0.0124 | 0.99 | 0.7715 | 0.995 | 0.5873 | 0.991 |
| Cumulative Sums | 0.4559 | 0.992 | 0.1917 | 0.989 | 0.1644 | 0.988 | 0.7238 | 0.983 | 0.9627 | 0.989 |
| Runs | 0.5626 | 0.989 | 0.5261 | 0.988 | 0.9216 | 0.991 | 0.1581 | 0.988 | 0.975 | 0.994 |
| Longest Run | 0.0032 | 0.984 | 0 | 0.925 | 0.0001 | 0.979 | 0.199 | 0.986 | 0.0262 | 0.997 |
| Rank | 0.1389 | 0.987 | 0.8219 | 0.991 | 0 | 0 | 0.9781 | 0.994 | 0.4673 | 0.99 |
| FFT | 0.4154 | 0.99 | 0.5042 | 0.986 | 0.1413 | 0.987 | 0.0338 | 0.987 | 0.6808 | 0.983 |
| Non-overlapping Template | 0.1672 | 0.983 | 0.4208 | 0.979 | 0.6392 | 0.979 | 0.9627 | 0.981 | 0.5382 | 0.981 |
| Overlapping Template | 0.5382 | 0.988 | 0 | 0.922 | 0.081 | 0.981 | 0.9549 | 0.986 | 0.4208 | 0.985 |
| Universal | 0.5362 | 0.99 | 0.9938 | 0.992 | 0.608 | 0.991 | 0.2925 | 0.992 | 0.4731 | 0.99 |
| Approximate Entropy | 0.7014 | 0.995 | 0 | 0 | 0.5022 | 0.99 | 0.0932 | 0.991 | 0.1856 | 0.984 |
| Random Excursions | 0.6983 | 0.9821 | 0.0052 | 0.9837 | 0.1837 | 0.9822 | 0.413 | 0.9802 | 0.98 | 0.9872 |
| Random Excursions Variant | 0.5827 | 0.9902 | 0.6559 | 0.9837 | 0.1756 | 0.9854 | 0.3115 | 0.9851 | 0.2124 | 0.9872 |
| Serial | 0 | 0 | 0.9357 | 0.99 | 0 | 0 | 0 | 0 | 0 | 0.041 |
| Linear Complexity | 0.2609 | 0.989 | 0.3804 | 0.987 | 0.874 | 0.991 | 0.0743 | 0.989 | 0.3821 | 0.993 |
| Number of removed blocks | 0 | | 0 | | 0 | | 0 | | 41 | |
| Result | Failure | | Failure | | Failure | | Failure | | Failure | |

Fig. A3: (color online). Results of the NIST test with small K values showing significant failure. Results of NIST Special Publication 800-22 statistical tests, using 1000 samples of 1 Mb data and significance level α = 0.01. The data was experimentally obtained using the scheme presented in fig. 1. A "success" result requires that the P value (uniformity of P values) should be larger than 0.0001 and the proportion should be greater than 0.980561 (for Random Excursions and Random Excursions Variant tests the proportion value might be even lower). The significant failure of the NIST test is evident by the failure of almost all blocks in one or several tests as also indicated by vanishing proportion and P values. Test results are shown (left to right): Embedding scheme 1 with K = 13 and G = 5 (fig. 2(b)); Embedding scheme 2 with K = 15, G = 5 m = 1 and q = 7 (fig. 2(c)); Embedding scheme 2 for a different sequence than the one presented in the manuscript with K = 13, G = 5 and a parity of multiple bits m = 1 and q = [2,6,9,12] (fig. 2(c)).

| Statistical test | Scheme 1 (sequence 1) K = 23, G = 5 | | Scheme 1 (sequence 1) K = 63, G = 5 | | Scheme 1 (sequence 1) K = 81, G = 5 | |
|---|---|---|---|---|---|---|
| | P value | Proportion | P value | Proportion | P value | Proportion |
| Frequency | 0.0062 | 0.991 | 0.3161 | 0.993 | 0.1019 | 0.992 |
| Block Frequency | 0.4827 | 0.993 | 0.5687 | 0.99 | 0.7811 | 0.99 |
| Cumulative Sums | 0.3425 | 0.988 | 0.7811 | 0.991 | 0.3071 | 0.987 |
| Runs | 0.2023 | 0.994 | 0.8378 | 0.992 | 0.3176 | 0.99 |
| Longest Run | 0.4635 | 0.994 | 0.0866 | 0.989 | 0.3753 | 0.99 |
| Rank | 0.6704 | 0.989 | 0.6496 | 0.992 | 0.4578 | 0.991 |
| FFT | 0.1886 | 0.991 | 0.01 | 0.991 | 0.8645 | 0.987 |
| Non-overlapping Template | 0.7055 | 0.982 | 0.9549 | 0.982 | 0.6059 | 0.983 |
| Overlapping Template | 0.8832 | 0.989 | 0.1917 | 0.989 | 0.9844 | 0.988 |
| Universal | 0.344 | 0.991 | 0.848 | 0.99 | 0.5102 | 0.983 |
| Approximate Entropy | 0.4083 | 0.993 | 0.7981 | 0.988 | 0.7177 | 0.989 |
| Random Excursions | 0.6303 | 0.9887 | 0.6678 | 0.9838 | 0.1739 | 0.9841 |
| Random Excursions Variant | 0.0535 | 0.9806 | 0.0346 | 0.9887 | 0.3518 | 0.9841 |
| Serial | 0.1876 | 0.991 | 0.232 | 0.988 | 0.6973 | 0.99 |
| Linear Complexity | 0.8693 | 0.989 | 0.7459 | 0.988 | 0.2689 | 0.991 |
| Number of removed blocks | 38 | | 24 | | 0 | |
| Result | Success | | Success | | Success | |

Fig. A4: (color online). Results of the NIST test for embedding random message in scheme 1. Results of NIST Special Publication 800-22 statistical tests, using 1000 samples of 1 Mb data (the same as in fig. 3) and significance level $\alpha = 0.01$. The random bit sequence was generated using the scheme presented in fig. 1. Embedding scheme 1 (fig. 2) was implemented where the last bit of each K-segment is the parity of its preceding K-1 bits plus a random bit as described in Eq. (3). A "success" result requires that the P value (uniformity of P values) should be larger than 0.0001 and the proportion should be greater than 0.980561 (for Random Excursions and Random Excursions Variant tests the proportion value might be even lower). Successful test results are shown (left to right): Embedding scheme 1 with K = 23 and G = 5; Embedding scheme 1 with K = 63 and G = 5; Scheme 1 with K = 81 and G = 5 (fig. 2(b)).

|  | Scheme 2 (sequence 1) K = 59, G = 5 m =1 , q = [7 18 31 43] | | Scheme 2 (sequence 1) K = 75, G = 5 m = 1 , q = [7 29 43 67] | |
|---|---|---|---|---|
| Statistical test | P value | Proportion | P value | Proportion |
| Frequency | 0.8786 | 0.991 | 0.5082 | 0.991 |
| Block Frequency | 0.4616 | 0.989 | 0.858 | 0.992 |
| Cumulative Sums | 0.7578 | 0.987 | 0.9549 | 0.99 |
| Runs | 0.7963 | 0.989 | 0.8273 | 0.985 |
| Longest Run | 0.7499 | 0.992 | 0.874 | 0.989 |
| Rank | 0.4466 | 0.982 | 0.5585 | 0.989 |
| FFT | 0.4924 | 0.987 | 0.8429 | 0.984 |
| Non-overlapping Template | 0.5503 | 0.981 | 0.7075 | 0.982 |
| Overlapping Template | 0.4808 | 0.984 | 0.4136 | 0.99 |
| Universal | 0.7753 | 0.991 | 0.6993 | 0.99 |
| Approximate Entropy | 0.9346 | 0.989 | 0.4866 | 0.988 |
| Random Excursions | 0.7084 | 0.9838 | 0.7258 | 0.986 |
| Random Excursions Variant | 0.6201 | 0.9822 | 0.9224 | 0.9844 |
| Serial | 0.1181 | 0.985 | 0.7055 | 0.988 |
| Linear Complexity | 0.4354 | 0.988 | 0.082 | 0.99 |
| Number of removed blocks | 21 | | 0 | |
| Result | Success | | Success | |

Fig. A5: (color online). Results of the NIST test for embedding scheme 2 with parity determined from multiple bits. Results of NIST Special Publication 800-22 statistical tests, using 1000 samples of 1 Mb data and significance level $\alpha = 0.01$. The random sequence was experimentally generated using the method described in fig. 1. Embedding scheme 2 was implemented where the last bit of each K-segment is the parity of a subset of 5 preceding bits. A "success" result requires that the P value (uniformity of P values) should be larger than 0.0001 and the proportion should be greater than 0.980561 (for Random Excursions and Random Excursions Variant tests the proportion value might be even lower). Successful test results are shown (left to right): Embedding scheme 2 with K = 59, G = 5 and parity of multiple bits m = 1 and q = [7,18,31,43] (fig. 2(c)); Embedding scheme 2 with K = 75, G = 5 and parity of multiple bits m = 1 and q = [7,29,43,67] (fig. 2(c)).

\* \* \*